\documentclass[11pt]{article}
\pdfoutput=1
\usepackage{amsmath}
\usepackage{amssymb} 
\usepackage{graphicx}
\usepackage{color}

\definecolor{cbl}{rgb}{0,0,1}                % bleu

\newcommand{\half}{\frac{1}{2}}
\newcommand{\bra}[1]{\langle #1 |} 
\newcommand{\ket}[1]{| #1 \rangle } 
\newcommand{\vev}[1]{\langle #1 \rangle}

\newcommand{\mM}{\mathfrak{M}}
\newcommand{\iI}{\mathfrak{I}}

\title{Real time imaging of quantum and thermal fluctuations:\\ the case of a
  two-level system.} 
\author{}
\date{}

\begin{document}
\maketitle
\pagestyle{empty}

\vskip -1.5 truecm 

\centerline{Michel Bauer ${}^{\spadesuit~}$ and Denis Bernard${}^{\clubsuit~}$}

\vskip 0.5 truecm

\noindent
\small{${}^\spadesuit$ Institut de Physique Th\'eorique de Saclay, CEA-Saclay $\&$ CNRS, 91191 Gif-sur-Yvette, France.}\\
\small{$^\clubsuit$ Laboratoire de Physique Th\'eorique de l'ENS, CNRS $\&$ Ecole Normale Sup\'erieure de Paris, France.}
\vskip 1.0 truecm

\pagestyle{plain}

\begin{abstract}

  A quantum system in contact with a heat bath undergoes quantum transitions
  between energy levels upon absorption or emission of energy quanta by the
  bath. These transitions remain virtual unless the energy of the system is
  measured repeatedly, even continuously in time. Isolating the two
  indispensable mechanisms in competition, we describe in a synthetic way the
  main physical features of thermally activated quantum jumps. Using classical
  tools of stochastic analysis, we compute in the case of a two-level system the
  complete statistics of jumps and transition times in the limit when the
  typical measurement time is small compared to the thermal relaxation time. The
  emerging picture is that quantum trajectories are similar to those of a
  classical particle in a noisy environment, subject to transitions \`a la
  Kramer in a multi-well landscape, but with a large multiplicative noise.
\end{abstract}

Keywords: Quantum jumps, Thermal fluctuations, Quantum trajectories, Stochastics
processes.

MSC: 60G17, 81S22, 82C31.

\vskip 0.5 truecm

\section{Introduction.}

Quantum jumps (Q-jumps) have been observed in strong resonance fluorescence
\cite{NaSaDe86,SaNeBlTo86} or in single atom \cite{Wine86} experiments. They are
abrupt system transitions from one quantum state to another. They were of course
known to the fathers of quantum mechanics, in particular in their realisation as
quantum state collapses during macroscopic measurements \cite{VonNeu}. Hundred
years ago Bohr \cite{Bohr} proposed that interaction of light and matter induces
transitions of an atom internal state with emission or absorption of a photon. A
more modern point of view, through the notion of {\it quantum trajectories} and
its Bayesian interpretation, is that these jumps reflect updatings of the
observer's knowledge and thus are not objective physical events independent of
the observer. Quantum trajectories \cite{Charm93,DC92,DC08} code for the
evolution of a quantum system under continuous observation or monitoring
\cite{WiseMil}. They are at the core of the quantum Monte Carlo method
\cite{DC92,DC08}, and they were recently observed in circuit QED \cite{Sid13}.
Without observations and measurements there would not be any jumps: these are
thus detector dependent \cite{WiGa12}, and in particular not instantaneous.

The aim of this letter is to extend the knowledge on quantum jump
dynamics, see e.g.  \cite{PleKni97} for a review, by analysing those
induced by thermal fluctuations. We aim at describing, theoretically and
analytically, what an observer who continuously measures the energy of a quantum
system in contact with a thermal bath is going to report. 

It is basic common knowledge -- say since Boltzmann -- that systems in contact
with a reservoir at fixed temperature undergo thermal fluctuations. These
fluctuations are induced by transitions from one energy level to another upon
absorption or emission of energy quanta by the thermal bath, in a way similar to
atom quantum jumps induced by photon emission or absorption. However, as for
Bohr's quantum jumps these transitions remain virtual as long as one does not
observe them. Measuring continuously the system's energy reveals them but at the
price that the system state becomes random with values depending on these
measurement outputs (whose probability distribution is induced by the quantum
mechanical rules for measurement).  Progresses on quantum non-demolition
measurements \cite{QND} recently lead to the observation of thermal fluctuations
in cavity QED \cite{Ha07}. These measurements were done by probing recursively
the QED cavity at very low temperature with series of Rydberg atoms. This is the
scheme that we shall adopt: we are going to describe the continuous observation
of a quantum system in contact with a reservoir by its interaction with series
of quantum probes subject to projective measurements. A continuous time limit of
this scheme leads to stochastic differential equations describing state
evolutions, called quantum trajectories, for systems under continuous time
measurement \cite{Bar,Bar2,Bel,Bel2}.  Our study of thermally activated quantum
jumps is based on analysing these stochastic differential equations but after
having minimised the number of inputs used in the description while keeping the
physics correct, see eqs.(\ref{Qtherm},\ref{Qeq}).  We shall analyse them in
some details in the case of a two state system, that is a Q-bit. Two relevant
time scales are involved: that associated to thermal relaxation, denoted
$\tau_{\rm therm}$, and that coding for the characteristic measurement time
$\tau_{\rm meas}$ -- measurement are not instantaneous.

For a Q-bit system, we show what is physically expected, namely:\\
$\bullet$ The thermal quantum trajectories possess an invariant measure which
concentrates on the Gibbs state when the measurement time $\tau_{\rm meas}$ goes
to zero, see eq.(\ref{limPstat}). The convergence in time toward this steady
state is exponential with a rate determined by the relaxation time $\tau_{\rm
  therm}$.\\
$\bullet$ For $\tau_{\rm meas}\ll \tau_{\rm therm}$, quantum trajectories jump
over and over between states which are asymptotically close to the energy
eigen-states. The mean waiting times between Q-jumps are of order $\tau_{\rm
  therm}$ and their ratio are given by Boltzmann factors, in accordance with
ergodicity, see eq.(\ref{tauwait}). This contrasts with the zero temperature
case for which these mean waiting times are of order $\tau_{\rm sys}^2/\tau_{\rm
  meas}$ with $\tau_{\rm sys}$ the characteristic time of the system Hamiltonian
evolution, in accordance with the quantum Zeno effect. We determine the full
statistics of the jump process, see eq.({\ref{Tdistri}) below.\\
  $\bullet$ The thermally induced quantum jumps are not instantaneous but their
  mean transition times are fixed by the measurement time up to logarithmic
  correction, that is $\tau_{\rm transit}\sim 4\tau_{\rm meas}\, \log(\tau_{\rm
    therm}/\tau_{\rm meas})$ for $\tau_{\rm meas}\ll \tau_{\rm therm}$, see
  eq.(\ref{Tjump}).  

  Weak measurements of Q-bits and their quantum trajectories have of course been
  analysed in the past, especially at zero temperature
  \cite{KorAr01,GoMil01,Gam08,Jacobs-et-al2007}. At zero temperature, quantum
  jumps arise from the competitive contributions of the Hamiltonian evolution
  and of the continuous measurement if the observable one is measuring does not
  commute with the Hamiltonian. In the thermal case the quantum jumps arise from
  the competition between the dissipative thermal evolution and the continuous
  measurement of the system's energy.  Our approach is based on mapping the
  quantum trajectory problem on that of a noisy particle (with only one degree
  of freedom) in a two well potential subject to thermally activated Kramer's
  transitions between the potential minima. This mapping is slightly different
  from the usual situation in the sense that quantum trajectory problems with
  small measurement times correspond to Kramer's problems with large noise, but
  multiplicative and of a particular form. We are nevertheless able to identify
  a Q-jump with a Kramer's like transition and to make a quantitative study of
  the statistics of jumps. To prove the above results we employ standard tools
  from probability theory, and we assume the reader to be familiar with those.

\section{Measurements and thermal quantum jumps.}

To visualise thermal fluctuations demands to measure continuously in time the
quantum system in contact with a thermal bath.  We need to describe both the
process induced by the continuous observation and that due to the interaction
with the bath at temperature $T$. A possible way for grasping what continuous
time measurement is consists in getting it from the continuous time limit of
repeated interactions, and this is the point of view that we shall adopt. So, we
consider a quantum system, called {\it the system}, recursively probed via
interaction with series of auxiliary identical quantum systems, called {\it the
  probes}. Von Neumann measurements are then implemented on the probes and this
series of indirect measurements is what constitutes the repeated, or continuous,
observation of the system. Being interested in thermal fluctuations we choose
the probes to indirectly measure the system's energy observable, alias the
system's Hamiltonian. We first describe the discrete setting and then its
continuous time scaling limit. For simplicity we take both the system Hilbert
space ${\cal H}_s$ and the probe Hilbert space ${\cal H}_p$ finite dimensional.
To be specific we choose ${\cal H}_p=\mathbb{C}^2$.

\underline{\it Discrete setting.}\\
\indent In the discrete setting, the system dynamics is a repeated alternation
of two dynamical maps : $i)$ that due to indirect measurement of the system's
energy, and $ii)$ that induced by interaction with the reservoir. We represent
the latter by a non-random completely positive quantum dynamical map,
\begin{eqnarray}\label{Qmap}
 \rho \to \sum_b B_b\, \rho\, B^\dag_b \quad {\rm with}\ \sum_b B^\dag_b
 B_b=\mathbb{I}, 
 \end{eqnarray}
 for some set of operators $B_b$ acting on the system Hilbert space. We don't
 need to specify them explicitly yet but we shall do it in the continuous time
 setting.

 Indirect measurement is modelled as follows, see e.g.\cite{BB11,BBB12}. At each
 time step an independent copy of the probe, prepared in a state $\ket{\psi}$,
 interacts with the system during a time duration $\delta$. Let $U_{\rm
   meas}=e^{-i\delta H_{\rm meas}}$ be the unitary operator acting on ${\cal
   H}_s\otimes {\cal H}_p$ coding for this interaction. After this interaction
 has taken place, a probe observable with eigen-vectors $\ket{i}\in {\cal H}_p$
 is projectively measured, giving some random output $i$ with value in the set
 of the observable eigen-values \footnote{For simplicity we assume that the
   probe observable which is measured has a non-degenerate spectrum.}. Because
 the system gets entangled with the probe during the interaction, the cycle of
 probe interaction and measurement induces a random evolution of the system
 density matrix $\rho$, called a quantum trajectory,
\begin{eqnarray} \label{Qmeas}
 \rho \to F_i\, \rho\, F^\dag_i/\pi_i,\quad {\rm with\ probability}\ \pi_i:={\rm
   Tr}(F_i\, \rho\, F^\dag_i), 
 \end{eqnarray}
 where $F_i:=\bra{i}U_{\rm meas}\ket{\psi}$. Unitarity of $U_{\rm meas}$ and
 completeness of the basis $\ket{i}$ imply $\sum_i F_i^\dag\, F_i= \mathbb{I}$
 and thus $\sum_i \pi_i=1$. That is, the $F_i$'s define a positive operator
 valued measure (POVM). These indirect measurements aim at measuring -- or at
 getting some information on -- a system observable ${\cal O}$, implying that
 the interaction evolution operator must be of the form $U_{\rm
   meas}=\sum_\alpha \ket{\alpha}\bra{\alpha}\otimes U_\alpha$ with
 $\ket{\alpha}\in {\cal H}_s$ the eigen-basis of ${\cal O}$ and $U_\alpha$
 unitary operators acting on ${\cal H}_p$. In such a case,
 $F_i=\ket{\alpha}\bra{i}U_\alpha\ket{\psi}\bra{\alpha}$ and the indirect
 measurement (\ref{Qmeas}) preserves the states $\ket{\alpha}$, as does a Von
 Neumann measurement of the observable ${\cal O}$. In the present situation we
 choose the observable to be the system's Hamiltonian.
 
 The process then consists in repeating alternatively the two evolutions
 (\ref{Qmeas},\ref{Qmap}). It is random because at each time step the probe
 measurement output is random with occurrence probabilities induced by quantum
 mechanical rules.

\underline{\it Continuous time  setting.}\\
\indent The discrete setting is closer to the actual physical realisation but
the continuous time setting can be analysed more thoroughly, and this is the one
we shall use. Taking the continuous time limit is a way to understand what are
the behaviours of the discrete evolution generated by (\ref{Qmap},\ref{Qmeas}).
It is valid when the time duration $\delta$ of each measurement cycle is much
shorter than any other time scale involved in the process. In practice one
takes the limit $\delta\to 0$ after a proper rescaling of the interaction
strength in $U_{\rm meas}$ to avoid the quantum Zeno effect. See e.g.
refs.\cite{Pel,BBB12}. The continuous time evolution equation for the system
density matrix in contact with a thermal bath and under continuous time
measurement is then of the form
\begin{eqnarray} \label{Qtherm} 
d\rho = (d\rho)_{\rm therm} + (d\rho)_{\rm meas},
\end{eqnarray}
where $(d\rho)_{\rm therm}$ is the thermal evolution and $(d\rho)_{\rm meas}$ is
the random evolution induced by the repeated indirect measurements.

The thermal evolution, which arises as the continuous time limit of the quantum
dynamical map (\ref{Qmap}), is described by a Lindblad equation \cite{Lind76},
\[ (d\rho)_{\rm therm}= L_{\rm therm}(\rho)\,dt .\] Recall that the Lindbladian
$L_{\rm therm}(\rho)$ is linear in $\rho$. We assume that without measurement
the system reaches the Gibbs steady state and this requires that $L_{\rm
  therm}(e^{-\beta H_{\rm sys}})=0$ with $\beta:=1/k_BT$ and $H_{\rm sys}$ the
system Hamiltonian. To be able to give a meaningful interpretation of the
quantum and thermal energy fluctuations we furthermore suppose that the flows
generated by the system Hamiltonian and the thermal Lindbladian commute, that is
$[H_{\rm sys}, L_{\rm therm}(\rho)]=L_{\rm therm}([H_{\rm sys},\rho])$ for any
system density matrix $\rho$.

The evolution $(d\rho)_{\rm meas}$ arises as the continuous time limit of the
probe interaction and measurement cycles. This evolution is random because the
probe measurement outputs are random. It is formulated in terms of stochastic
equations called either stochastic master equations (SME) \cite{Bar,Bar2} or
Belavkin's equations \cite{Bel,Bel2}. We restrict ourselves to diffusive cases
which correspond to situations in which the state of the probe before
interaction has non zero overlap with any eigen-state of the probe observable
${\cal O}$, i.e. in our notation $\bra{i}\psi\rangle\not=0$ for any $i$. For
spin half probes, Belavkin's equation is then of the form
\[ (d\rho)_{\rm meas}=L_{\rm meas}(\rho)\,dt + D_{\rm meas}(\rho)\, dB_t\] where
$B_t$ is a Brownian motion, $(dB_t)^2=dt$, echoing in the continuous time
scaling limit the statistics of probe measurement outputs\footnote{Recall that
  for spin half probes the measurement outputs are strings of plus or minus, say
  $(++-+--\cdots)$, in one-to-one correspondence with classical random walks on
  the line, so that their continuous time limit are naturally related to one
  dimensional Brownian motion.}, with specific Lindbladian $L_{\rm meas}$ and
non-linear diffusion coefficient $D_{\rm meas}$. We shall make them explicit in
the two level case below, but see refs.\cite{Bar,Bar2,Bel,Bel2,Pel,BBB12} for a
detailed description in the general case.  Let us however note that $L_{\rm
  meas}(\rho)$ is linear in $\rho$ while $D_{\rm meas}(\rho)$ is quadratic in
$\rho$. Also, taking the continuous limit of discrete probe measurements, one is
lead naturally to stochastic equations written in the It\^o formalism. So we
stick to the It\^o convention in the rest of this article, even though this is
by no means mandatory and usual modifications could be used to switch for
instance to the Stratanovich convention. See refs.\cite{Pel,BBB12} for a
derivation of these equations from the discrete time formulation.

\underline{\it The two level case.}\\
Let us specialise to a Q-bit system analysed with spin half probes, the case we
want to study here in some detail. Let $\ket{0}$ and $\ket{1}$ be the system
energy eigen-states, with energy $0$ and $\omega$ respectively $(\omega>0)$. As
long as one is only interested in properties related to the energy observable,
one only needs to know the time evolution of the diagonal matrix elements of the
system density matrix. If the flows generated by the system Hamiltonian and the
thermal Lindbladian commute the evolution of these elements is independent of
that of the off-diagonal elements. So we may safely restrict ourselves to
diagonal system density matrices,
\[ \rho = Q\, \ket{0}\bra{0}+ (1-Q)\, \ket{1}\bra{1},\] with $Q$ the probability
for the system to be in the ground state $\ket{0}$. If one is continuously
measuring the energy, the time evolution (\ref{Qtherm}) reduces to the following
quantum trajectory equation for $Q$:
\begin{eqnarray} \label{Qeq}
dQ_t=(dQ)_{\rm therm} + (dQ)_{\rm meas} =\lambda\,[p-Q_t]\, dt+ \gamma\,
Q_t(1-Q_t)\, dB_t, 
\end{eqnarray}
with $p$ the probability to be in the ground state at thermal equilibrium, $p:=
{1}/{(1+e^{-\beta\omega})}$. For $\beta=0$ we have $p=1/2$. Here $B_t$ is a
normalised Brownian motion related to the continuous time limit of the probe
measurement outputs. Eq.(\ref{Qeq}) involves two time scales:
$\tau_{\rm therm}:=\lambda^{-1}$  the thermal relaxation time, and $\tau_{\rm
  meas}:=\gamma^{-2}$ the measurement time. We define the dimensionless ratio
$\sigma:= 2\,{\tau_{\rm meas}}/{\tau_{\rm therm}}\ll 1$ and shall often assume
$\sigma \ll  1$, i.e. $\tau_{\rm meas}\ll \tau_{\rm therm}$.

Eq.(\ref{Qeq}) has a simple interpretation and may be derived simply on the
basis of symmetry arguments. The first thermal term only contains a drift term
-- no noise --, and is linear in $Q$ as are any Lindbladian evolutions, and it
vanishes at thermal equilibrium.  The second term is quadratic in $Q$ as are
Belavkin's equations for continuous time measurement, it only involves a noisy
term -- no drift -- because $Q$ has to be a martingale \cite{BB11} if the
observable one is measuring is the energy and it vanishes for $Q=0$ and $Q=1$
corresponding to the two energy eigen-states.  If one prefers, one may derive
eq.(\ref{Qeq}) from the explicit form of the Lindblad operators. The thermal
Linbladian reads:
\[ L_{\rm therm}(\rho)=-i\frac{\omega}{2}[\sigma_z,\rho]+\lambda
p\big(\sigma_-\rho\sigma_+-\frac{1}{2}\{\sigma_+\sigma_-,\rho\}\big) +
\lambda(1-p)\big(\sigma_+\rho\sigma_--\frac{1}{2}
\{\sigma_-\sigma_+,\rho\}\big),\] 
and this is the most general Lindbladian whose flow commutes with that generated
by $\sigma_z$. Here $\sigma_{z,+,-}$ are the usual Pauli matrices in the basis
$\ket{0},\, \ket{1}$, and $[\cdot,\cdot]$ and $\{\cdot,\cdot\}$ denote the
commutator and the anti-commutator respectively.  The Lindbladian $L_{\rm meas}$
and diffusion $D_{\rm meas}$ operators associated to the continuous measurement
of $\sigma_z$ can be written as
\[ L_{\rm meas}(\rho)= -\frac{\gamma^2}{32} [\sigma_z,[\sigma_z,\rho]],\quad
{\rm and}\quad D_{\rm meas}(\rho)=\frac{\gamma}{4}\big( \{\sigma_z,\rho\} -
2\rho\, {\rm tr}(\rho\sigma_z)\big).\] As it should be, $L_{\rm meas}(\rho)=0$
for $\rho=\frac{1}{2}\mathbb{I}+ (Q-\frac{1}{2})\sigma_z$ diagonal as above.
 
\begin{figure} 
  \centerline{\mbox{\includegraphics[width=1.2\textwidth]{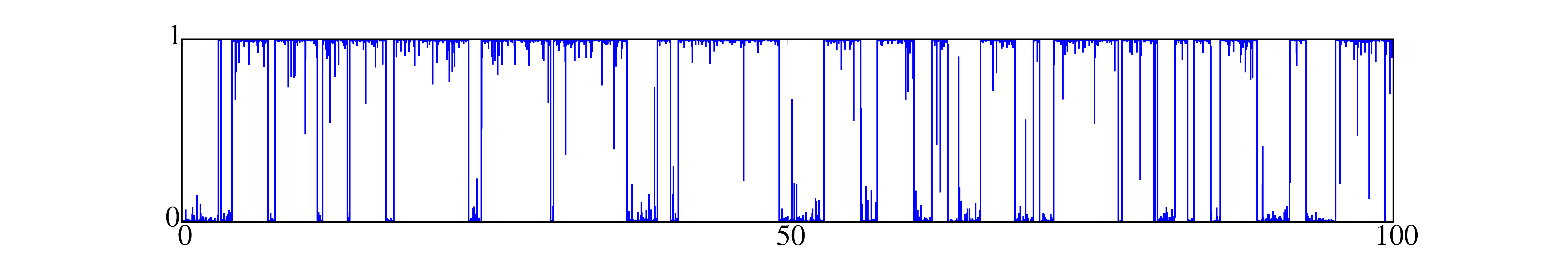}}} 
  \vskip -.6 truecm \caption{\small{\it Thermally activated quantum jumps: A
      sample realisation for $p=.6$, $\tau_{\rm therm}=1$,and $\tau_{\rm
        meas}=10^{-6}$ of the $Q_t$ trajectories for a Q-bit in contact with a
      thermal bath and under continuous energy measurement. Note the structure
      of the unsuccessful excursions in between jumps. }}
\label{fig:continuous-jump}
\end{figure}

A sample solution of the discrete version of (\ref{Qeq}) is shown in
{Fig.\ref{fig:continuous-jump}}. It clearly exhibits thermally activated quantum
jumps between the two energy eigen-states. The aim of the following is to give a
description of the dynamics and statistics of these jumps. They arise from the
competitive contributions of the evolution and the continuous measurement. Their
statistics will be derived by analysing the random quantum trajectories
(\ref{Qeq}) of $Q$. They resemble Kramer's like transitions but not quite
because we are dealing with $\gamma^2\gg \lambda$ which is not the usual small
noise limit \cite{Small,FW}.  However, (\ref{Qeq}) shows that the noise becomes
negligible for $Q$ close to $0$ (precisely $Q \ll \sigma^{1/2}$) or close to
$1$, and this is the reason why the small $\sigma$ limit is manageable. Below we
describe the invariant measure for (\ref{Qeq}), the waiting time and transition
time statistics of (\ref{Qeq}).

Before entering into the detailed description, let us deal with a question of
interpretation. Density matrices code for ensemble averages, i.e. they define
measures (in the sense of probability) to compute expectations of system
observables when experiments are repeated under identical conditions, say
identical quantum systems in contact with reservoirs. When implementing Von
Neumann measurements on the probes, we are still working with ensemble averages
for the system plus reservoir even though we have got specific outputs for these
measurements. The meaning of the quantum trajectories (\ref{Qeq}) is that they
code for the measure (in the sense of probability) for system observables
conditioned on having got some given outputs for the probe measurements (that
is, a measure on ensemble of systems identically prepared and all having got
these probe outputs). Furthermore, any given specific realisation of the system
plus reservoir evolution is unitary (and this is the point of view associated to
quantum stochastic differential equation \cite{qSDE}) but the reservoir is
macroscopic and usually not observed nor controlled. What is expected, and
usually assumed, is that there is some kind of concentration of measure so that
there is no relevant rare event. And hence, predictions based on using density
matrices to compute (conditional) ensemble averages are believed, or expected,
to faithfully represent typical behaviours as well as the mean behaviour.

\section{Statistics of jumps}

We now present results on the statistics of jumps, including properties of the
invariant measure and its behaviour when $\sigma\to0$, a description of the
statistics of the jump process especially in the limit $\sigma\to 0$. We choose
to organise the presentation by first giving a -- somewhat informal --
description of these properties and then elements of proof -- including more
precise statements.

\underline{\it The invariant measure.}\\
The quantum trajectories (\ref{Qeq}) admit an invariant measure (in the sense of
probability) which possesses two peaks respectively centred close to $0$ and $1$
(more precisely around $Q_-\simeq p\sigma$ and $Q_+\simeq 1-(1-p)\sigma$). It
slightly differs from the expected Gibbs measure for $\sigma$ finite
($\sigma:=2\tau_{\rm meas}/\tau_{\rm therm}$), but it converges in the limit of
vanishing $\sigma$ (i.e. for a vanishingly small measurement time),
\begin{eqnarray} \label{limPstat}
\lim_{\sigma\to0} dP_{\rm stat} = \big( (1-p) \delta(Q) + p\delta(Q-1) \big) dQ.
\end{eqnarray}
For such a simple system, any initial measure converges to the stationary
measure in the long run, so that $\lim_{t\to\infty}\mathbb{E}[ F(Q_t)] =
\mathbb{E}_{\rm stat}[ F(Q)]$ for any reasonable function $F$. In particular
$\mathbb{E}_{\rm stat}[Q]=p$ and $\lim_{t\to\infty}\mathbb{E}[\rho_t]= p\,
\ket{0}\bra{0}+ (1-p)\, \ket{1}\bra{1}$ is the Gibbs state $\propto e^{-\beta
  H_{\rm sys}}$. In absence of measurements, the thermalization time scale is
$\tau_{\rm therm}=\lambda^{-1}$. For instance
$\mathbb{E}[(p-Q_t)]=(p-Q_0)e^{-\lambda t}$. We prove that this relaxation time
is unmodified by the measurement process, $\lim_{\sigma\to0}\,\tau_{\rm relax}=
\lambda^{-1}$.

{\it Sketch of the proof:} 
Let us
write (\ref{Qeq}) in the form $dQ_t=\gamma^2\,f(Q_t)dt + \gamma\, g(Q_t)dB_t$
and define a function $h(Q)$ by $\partial_Qh(Q)=-f(Q)/g^2(Q)$. By a classical
result (see e.g. \cite{ito-mckean1991}) the
invariant measure reads $dP_{\rm stat} = g^{-2}(Q)\, e^{-2h(Q)}\, dQ$.
Explicitely,
\[dP_{\rm stat}= \frac{1}{Z_\sigma}\,
\frac{dQ}{Q^{2-a_\sigma}(1-Q)^{2+a_\sigma}}\, \exp\big[
-{\sigma}\big(\frac{p}{Q}+\frac{1-p}{1-Q}\big)\big].\] with $Z_\sigma$ a
normalisation factor and $a_\sigma:=\sigma({2p-1})$. We infer that the density
$dP_{\rm stat}/dQ$ exhibits two peaks at $0 < Q_- < Q_+ <1$ and a minimum in
between, with $Q_-\simeq p\sigma$, and $1-Q_+\simeq (1-p)\sigma$ for
$\sigma\ll1$.
 
It is instructive to change variables, bringing (\ref{Qeq}) in a standard
normalized form. % Let $X:=\log(\frac{Q}{1-Q})$, so that $dQ=Q(1-Q)dX$.
The process $X_t:=\log(\frac{Q_t}{1-Q_t})$ takes value on the real axis since
$0<Q_t<1$. Using It\^{o}'s formula, (\ref{Qeq}) becomes $dX_t= -\gamma^2\,
V'(X_t)\, dt + \gamma\, dB_t$ with effective double-well potential
\[ V(X)= \frac{1}{2}(1-a_{\sigma})\log Q+ \frac{1}{2}(1+a_{\sigma})\log(1-Q) +
\frac{\sigma}{2}\big[ \frac{p}{Q} + \frac{1-p}{1-Q}\big].\] The height of the
barrier is logarithmic in $\sigma$. The invariant measure reads $dP_{\rm stat}
\propto e^{-2V(X)}\, dX$.

Eq.(\ref{limPstat}) may be proved by
computing the weight of $dP_{\rm stat}$ on the interval $[0,\alpha]$ for
$0<\alpha<1$, that is $\int_0^\alpha dP_{\rm stat}$ when $\sigma\to 0$. For
$\sigma$ small and $Q$ close to $0$, we may set $\sigma=0$ except in the
exponential term $(\frac{\sigma p}{Q})$ so that
\[ \int_0^\alpha dP_{\rm stat}\underset{\sigma\to0}{\sim} \frac{1}{Z_\sigma}
\int_0^\alpha \frac{dQ}{Q^2} e^{-\sigma p/Q}\underset{\sigma\to0}{\sim}
\frac{1}{Z_\sigma}(\frac{1}{ \sigma p}).\] The crucial point is that this
dominant contribution is $\alpha$-independent.  Similarly, computing the weight
of the measure on $[1-\alpha,1]$, we get $\int_{1-\alpha}^1 dP_{\rm stat}\,
\underset{\sigma\to0}{\sim} {1}/{(Z_\sigma \sigma}(1-p))$ which is also
$\alpha$-independent. This means that in the limit $\sigma \to 0$ all
the weight is concentrated at $0$ and $1$. The relative weight of the two peaks
is $(1-p)/p$ and this proves (\ref{limPstat}).

For $\sigma$ small we get the following estimates, approximating the Dirac point
measures
\begin{eqnarray*}
  dP_{\rm stat} \underset{\sigma\to0}{\sim} \begin{cases} \sigma p(1-p)\,Q^{-2}dQ\,\exp{(-{\sigma}\frac{p}{Q})}&  {\rm for}\ Q\to 0,\\ ~&~\\
    \sigma p(1-p)\,(1-Q)^{-2}dQ\, \exp{(-{\sigma}\frac{1-p}{1-Q})},& {\rm for}\ Q\to 1. \end{cases}
\end{eqnarray*}

The approach to the invariant measure is governed by the Fokker-Planck operator
${\cal H}_{\rm HP}$ associated to (\ref{Qeq}), and more precisely by its first
non zero eigen-value. As usual, this operator is not symmetric but it is
self-transposed up to a conjugation, ${\cal H}_{\rm FP}^T= e^{-2V}\, {\cal
  H}_{\rm FP}\, e^{+2V}$. The operator ${\cal H}_{\rm FP}^T$ is the operator
which codes for the evolution of expectations, that is $d\mathbb{E}[F(Q_t)]=\mathbb{E}[({\cal H}_{\rm FP}^T\cdot F)(Q_t)]dt$ for any function $F$. We know two eigen-functions of ${\cal H}_{\rm FP}^T$ which are $1$ and $p-Q$ with respective eigen-value $0$ and $-\lambda$, because $\mathbb{E}[p-Q_t]=(p-Q_0)e^{-\lambda t}$. Hence we know two eigen-vectors of ${\cal H}_{\rm FP}$: the stationary measure $P_{\rm stat} \propto e^{-2V}$ with zero eigen-value and $P_1\propto (p-Q)\, e^{-2V}$ with eigen-value $-\lambda$. These are the two first eigen-values because $P_{\rm stat}$ has no zero and $P_1$ a unique zero. Hence $\tau_{\rm relax}=\lambda^{-1}$, and the relaxation time is not modified by the measurement process.

\underline{\it The statistics of waiting times and jumps.}\\
This concerns the statistics of the time spent by the quantum trajectories near
the value $Q_-\simeq 0$ and $Q_+\simeq 1$ respectively, so that the Q-bit is
effectively close to the state $\ket{1}$ or $\ket{0}$. We prove that the limits
of the mean time 
$T_0$ (resp. $T_1$) the quantum trajectories spend near the state $\ket{0}$ (resp. $\ket{1}$) are
\begin{eqnarray} \label{tauwait}
\lim_{\sigma\to0}\,T_0= \tau_{\rm therm}/(1-p),\quad 
\lim_{\sigma\to0}\,T_1= \tau_{\rm therm}/p,
\end{eqnarray}
so that $\lim_{\sigma\to0}\,T_1/T_0= e^{-\beta\omega}$ as expected from ergodicity. Contrary
to the zero temperature quantum jumps, the mean waiting times of the thermally
activated jumps stay finite when the measurement time decreases, as they
should. The precise distribution is determined below in the limit of $\sigma\to 0$, 
see eq.(\ref{Tdistri}): starting at $Q_i$, the distribution of the first passage time at $Q_f$
($0 \leq Q_i < Q_f \leq 1$) is a mixture of a Dirac peak at $0$ and an
exponential distribution whose parameter depends on $Q_f$. For $Q_i=0$ the
distribution is purely exponential. The succession of passage times at $0$ and
$1$ is thus kind of (and exactly for $p=1/2$) a Poisson point process.

{\it Sketch of the proof:} 
We define these waiting times as the times needed to transit from one potential
minimum to the other\footnote{Recall that for $Q\simeq 0$ (resp. $Q\simeq 1$)
  the state is close to $\ket{1}$ (resp. $\ket{0}$).}. To be specific we define
$T_1$ (resp. $T_0$) as the first instance the quantum trajectory $Q_t$, starting
at $Q_i$ close to $0^+$ (resp. $1^-$) reaches $Q_f$ close to $1^-$ (resp.
$0^+$). These are stopping times, and to compute their mean is a standard
problem in stochastic differential equations, see e.g. \cite{Oksen,Kal}. Let us
write (\ref{Qeq}) in the form $dQ_t=\gamma^2\,f(Q_t)dt + \gamma\, g(Q_t)dB_t$
and define a function $h(Q)$ by $\partial_Qh(Q)=-f(Q)/g^2(Q)$, so that the
invariant measure reads $dP_{\rm stat} = g^{-2}(Q)\, e^{-2h(Q)}\, dQ$. By a
classical formula, the mean time spend around $Q_i$ (close to the left minimum
of the potential) is then
\[ T_1=2\gamma^{-2} \int_{Q_i}^{Q_f} dQ\,e^{2h(Q)}\int_0^Q dP_{\rm stat}.\]
This formula is in fact a limiting formula for the time to leave an interval :
in our case, the interval is $[0,Q_f]$ and the singularity at $0$ ensures that
the exit is always at $Q_f$. This explains the presence of $0$ as the lower
bound in the integral over $dP_{\rm stat}$. We do this integral as above when proving convergence of the stationary measure and approximate $e^{2h(Q)}$ by $e^{\sigma p/Q}$ for $\sigma$ small enough. This yields
\[ T_1\underset{\sigma\to0}{\sim} 2\gamma^{-2} \int_{Q_i}^{Q_f} dQ\, e^{\sigma p /Q}\int_0^Q \frac{dQ'}{Q'^2}e^{-\sigma{p}/{Q'}}
\underset{\sigma\to0}{\sim} \frac{2(Q_f-Q_i)}{\sigma\gamma^2}\frac{1}{p} .\]
This formula is valid for any $Q_i < Q_f$, and quantum jumps correspond to
$Q_i\simeq 0$ and $Q_f\simeq 1$, leading to $T_1 \underset{\sigma\to0}{\sim}
\frac{2}{\sigma\gamma^2}\frac{1}{p}$ for $\sigma$ small.
The mean time $T_0$ spend around $Q_+\simeq 1$ is simply obtained from the previous integral with $p\to 1-p$. Recall that $\sigma=2\tau_{\rm meas}/\tau_{\rm therm}$ and $\tau_{\rm meas}=\gamma^{-2}$. This proves (\ref{tauwait}).

To deal with the full distribution of waiting times, we observe that, by the
strong Markov property, we have 
\begin{equation}
{\mathbb E}[e^{-uT_{i\to f}/\tau_\text{therm}}]= e^{-\int_{Q_i}^{Q_f} \varphi(Q,u) dQ}.
\end{equation}
for some function $\varphi(Q,u)$, where $T_{i\to f}$ is by definition the random
time it takes to go from $Q_i$ to $Q_f > Q_i$, so $T_{i\to f}$ is a stopping
time.  We use a standard martingale trick (see e.g. chapter 7 \cite{Kal}): by
construction the conditional expectation $N_t:={\mathbb E}[e^{-uT_{i\to
    f}/\tau_\text{therm}}|\mathcal{F}_t]$ is a martingale. On the other hand, by
the Markov property, $N_t=e^{-ut/\tau_\text{therm}}e^{-\int_{Q_t}^{Q_f}
  \varphi(Q,u) dQ}$.  It\^o's formula applied to $N_t$ yields that
\[\sigma \big(-u+(p-Q)\varphi\big) +Q^2(1-Q)^2 \big(\varphi' +\varphi^2\big)=0,\]
where $\varphi'$ is the derivative of $\varphi$ with respect to $Q$. The
shortest route to take the $\sigma \to 0$ limit is as follows\footnote{This
  relies on an interchange of limits that is easy to justify}. For
$\sigma=0$ the equation degenerates to $\varphi' +\varphi^2=0$, so that in this
limit $\varphi(Q,u)=\frac{\varphi_0(u)}{1+Q \varphi_0(u)}$ for some integration
``constant'' $\varphi_0(u)$. For small $Q$ the noise becomes irrelevant, so that
$T_{i\to f}$ becomes deterministic when both $Q_i$ and $Q_f$ are close to $0$.
In this regime, (\ref{Qeq}) degenerates to $dQ_t=\lambda p dt$, a motion at
constant speed. This gives $\varphi_0(u)=u/p $, and
$\varphi(Q,u)=\frac{u}{p+uQ}$.  Finally, in the limit $\sigma \to 0$
\[ \lim_{\sigma\to0}\, {\mathbb E}[e^{-uT_{i\to f}/\tau_\text{therm}}]= \frac{uQ_i+p}{uQ_f+p}.\]
Undoing the Laplace transform, we obtain that, for any Borel subset $B$ in
$\mathbb{R}$, 
\begin{equation} \label{Tdistri}
\lim_{\sigma\to0}\, {\mathbb P}[T_{i\to f}/\tau_\text{therm} \in B]= \frac{Q_i}{Q_f}\, {\mathbb
  I}_{\, 0 \in B} +\Big(1-\frac{Q_i}{Q_f}\Big)\frac{p}{Q_f} \int_B
e^{-s\frac{p}{Q_f}} ds, 
\end{equation} 
i.e.  the law of $T_{i\to f}$ is a mixture of a Dirac
peak at $0$ (weight $Q_i/Q_f$) and an exponential distribution of parameter
$\frac{p}{Q_f}$ (weight $1- Q_i/Q_f$). Intuitively when $\sigma \to 0$ but $Q_i
>0$ the trajectory starting at $Q_i$ has a chance to reach $Q_f$ without being
trapped in the potential well, hence the presence of a Dirac peak. However if
$Q_i \lesssim \sigma \to 0$, the trajectory has to go through the potential well
and waits there an exponential time. The $Q_f$-dependence of this exponential
time, already visible in the mean waiting time, shows that the situation is not
exactly covered by standard Kramer's arguments, i.e. it is not only the escape
from a potential well that counts. In the $Q$-coordinate, this is interpreted as
a multiplicative noise effect: when $Q(1-Q)$ departs significantly from $0$, the
noise becomes very large (a look at {Fig.\ref{fig:continuous-jump}} and its many
spikes may be illuminating at that point). In this $X$-coordinate, this is
interpreted as the fact that the potential wells are only logarithmically deep,
and their separation is very large.

\underline{\it The transition time statistics.}\\
To get a clue on the jump dynamics let us now look at the mean time needed to transit between the two energy eigen-states. These are independent of the direction of the transition, either from $Q\simeq 0$ to $Q\simeq 1$ or the reverse. We shall prove that they are determined by the measurement time up to logarithmic corrections (which may be large),
\begin{eqnarray} \label{Tjump}
 \tilde{\tau} \ \underset{\sigma\to 0}{\sim}\ 4\, \tau_{\rm meas}\ \log(\tau_{\rm therm}/\tau_{\rm meas}).
 \end{eqnarray}
This shows that the inside jump dynamics is dominated by the measurement procedure but also influenced by the other dynamical processes which are the sources of the fluctuations.

{\it Sketch of the proof:} 
The transition time is the time it takes to cross the barrier in events where
this is really what happens: the process is to start at one point $Q_i$ (on one
side of the barrier) and leave it for good until it reaches a point $Q_f$ (on
the other side of the barrier). This means we want to make statistics only on
certain events, i.e. we have to condition. If $Q_i$ is not a singular point for
the diffusion, a typical sample starting at $Q_i$ will, with probability $1$,
visit $Q_i$ again uncountably many times in an arbitrary small time interval.
This means that some limiting procedure is needed to define conditioning: one
starts the process at an intermediate point $Q$ between $Q_i$ and $Q_f$ and
conditions on the event that $Q_f$ is reached before $Q_i$, then the limit $Q
\to Q_i$ is taken. As we shall recall below, the equation for the conditioned
process does not depend on $Q_f$, but for the conditioned process the
point $Q_i$ is singular, so that even if the conditioned process is started at
$Q_i$, $Q_i$ is immediately left for good.

We begin by recalling some general formul\ae . A general reference completing
\cite{Oksen,Kal} for this discussion is \cite{ito-mckean1991}. If
$dQ_t=\gamma^2f(Q_t)dt +\gamma g(Q_t)dB_t$ is the diffusion equation describing
the time evolution of $Q_t$, we define the so-called scale function $s(Q)$ by
$s''/s'=-2f/g^2$. From the definition, the scale function is defined modulo an
affine transformation. The scale function is related to the previously
introduced function $h$ by the simple relation $s'(Q) \propto e^{2h(Q)}$ but for
the purpose of the present discussion, $s$ is slightly more convenient. It\^o's
formula shows that $s(Q_t)$ is a continuous (local) martingale, i.e. a
time-changed Brownian motion. A classical result, easily retrieved by standard
martingale techniques for instance, is that $\Pi_{[Q_i,Q_f]}(Q)$, the
probability starting at $Q \in [Q_i,Q_f]$ to exit the interval at $Q_f$ is
\[\Pi_{[Q_i,Q_f]}(Q)=\frac{s(Q)-s(Q_i)}{s(Q_f)-s(Q_i)}.\]
In fact, this is the origin of the scaling function: if $s(Q)$ is used as a new
``space'' variable, exit probabilities look like those of Brownian motion. This
is to be expected because exit probabilities do not involve a time
parameterisation, so they are the same for Brownian motion and time-changed
Brownian motion. The probability $\Pi_{[Q_i,Q_f]}(Q)$, abbreviated as $\Pi(Q)$
in the sequel, is relevant because we want to condition precisely on those
samples that contribute to it. This is achieved by a Girsanov transformation:
Girsanov's theorem states that the equation governing the initial motion
conditioned to exit $[Q_i,Q_f]$ at $Q_f$ is $dQ_t=\gamma^2\tilde{f}(Q_t)dt
+\gamma g(Q_t)d\tilde{B}_t$ where $\tilde{B}_t$ is a Brownian motion and
$\tilde{f}:=f+g^2 \frac{\Pi'}{\Pi}$. To say things in a different but equivalent
way: under conditioning, the process $B_t$ is not a Brownian motion anymore, but
$\tilde{B}_t:=B_t-\gamma \int_0^t g^2(Q_s)\frac{\Pi'(Q_s)}{\Pi(Q_s)}ds$ becomes
one. Note that $Q_f$ does not appear in
$\frac{\Pi'(Q)}{\Pi(Q)}=\frac{s'(Q)}{s(Q)-s(Q_i)}$ which has a simple pole with
residue $1$ at $Q=Q_i$, a singularity characteristic of conditioning not to pass
at $Q_i$.  This singularity implies that the transition time $\tilde \tau$ can be written as a
double integral: defining $\tilde{h}$ (the analogue of $h$ but for the conditioned
equation) by $\tilde{h}':=-\tilde{f}/g^2$, we get
\[ \tilde \tau=2\gamma^{-2} \int_{Q_i}^{Q_f} dQ\,e^{2\tilde{h}(Q)}\int_{Q_i}^Q
dq\, e^{-2\tilde{h}(q)}/g(q)^2.\] Note that in the inner integral the lower
bound is again the position of repelling singularity, i.e.  $Q_i$ for the
conditioned equation, just as it was $0$ for the initial equation.  Re-expressing
$\tilde{h}$ in terms of the scale function, the fact that the singularity comes
from conditioning leads to further simplifications via integrations by part,
leading finally to
\[ \tilde \tau= \frac{2\gamma^{-2}}{s(Q_f)-s(Q_i)} \int_{Q_i}^{Q_f} dQ \frac{ \big(
  s(Q)-s(Q_i)\big)\big(s(Q_f)-s(Q)\big)}{s'(Q) g^2(Q)}.\] These are general
formul\ae\ that we gave because the tricks are simpler to disentangle in the
general setting. It is reassuring to observe that the results for the exit
probabilities or for the transition time are indeed invariant under affine
transformations of the scale function.

In the case at hand, $f(Q)=\sigma(p-Q)/2$ and $g(Q)=Q(1-Q)$ leading to the
formula $s'(q) = \left(\frac{1-q}{q}\right)^{\sigma (2p-1)} e^{ \sigma
  (\frac{p}{q}+\frac{1-p}{1-q})}$. In the limit when $\sigma \to 0$ for
$Q$ staying in a $\sigma$ independant interval, $s(Q)$ becomes an affine function of $Q$, so that for fixed $Q_i$
and $Q_f$, the mean transition time $\tilde\tau$ is given by the integral
$2\gamma^{-2} \int_{Q_i}^{Q_f} \frac{dQ}{g^2(Q)}
\frac{(Q-Q_i)(Q_f-Q)}{(Q_f-Q_i)}$, and we get
\[\tilde \tau \underset{\sigma\to0}{\sim}\frac{2}{\gamma^2}\Big(
\frac{Q_i+Q_f-2Q_iQ_f}{Q_f-Q_i}\log \frac{Q_f(1-Q_i)}{Q_i(1-Q_f)}-2\Big).\] Note
that this formula is $p$-independent.

Up to now, we have taken $Q_i$ and $Q_f$ fixed and taken the limit $\sigma \to
0$. What we really want is more subtle, because we want $Q_i$ and $Q_f$ to sit
at the bottoms of the potential wells, which are $\sigma$-dependent. The trouble
is that near the bottoms, the scale function $s(Q)$ is not well-described by an
affine function. Ignoring this problem for a while, let us see what the na\"ive
limiting approach ``take the small $\sigma$ limit for fixed $Q_i$ and $Q_f$ and
then take $Q_i$ and $1-Q_f$ of order $\sigma$'' predicts.  Locating the
positions of the bottoms of the potential wells for small $\sigma$, we get $Q_i
\simeq p \sigma$ and $Q_f \simeq 1-(1-p)\sigma$. Plugging these asymptotics
blindly in the above formula for $\tilde \tau$ we get, for the transition time
from one potential well to the other in the small $\sigma$ limit,
\[\tilde \tau \underset{\sigma\to 0}{\sim}-\frac{4}{\gamma^2}\log \sigma
\underset{\sigma\to 0}{\sim} 4\, \tau_{\rm meas} \log(\tau_{\rm therm}/\tau_{\rm
  meas})\] as announced above. The rigorous justification of this formula,
taking properly into account the region where the affine approximation for
$s(Q)$ fails, requires some effort. The details are given in Appendix
\ref{app:details} using an argument suggested to us by an anonymous referee, see
especially eq.(\ref{eq:final-result}).  There is no dependence in $p$ in the
expression for $\tilde \tau$, but it is valid only for $\sigma \ll p(1-p)$.

\section{Non-sensitivity to initial conditions}

This very short section is less rigorous than the previous ones.  We give a
heuristic argument that for given probe measurement outcomes (i.e. for a given
realisation of the Brownian motion), two trajectories with different initial
conditions get closer at an exponential rate on a time scale of order $\tau_{\rm
  meas}$. This means that there is exponential memory loss. This generalises in
the presence of thermal noise the result of \cite{BB11} on repeated non-demolition
measurements.

The stochastic differential equation (\ref{Qeq}) only involves smooth driving
functions and the solutions of interest remain bounded in $[0,1]$. Thus its
solutions behave ``nicely'': two solutions with different initial conditions $0
\leq Q_0< Q'_0\le 1$ but for the same sample of Brownian motion cannot stick
together at any later time (i.e. strong existence and uniqueness of the solution
for a given realisation of the Brownian motion hold). As trajectories are
continuous, this means that $Q_t < Q'_t$ for every $t$.  For $\sigma$ small
enough, this implies that these two solutions cannot avoid getting very close
because both $Q_t$ and $Q'_t$ have to jump between values close to $0$ and $1$.
Once they are close, we can accurately linearise (\ref{Qeq}) around the solution
$Q_t$.

\underline{\it Linear theory.}\\ 
Keeping only first order terms in the deviation $\Delta_t:= Q'_t-Q_t$, we get
from (\ref{Qeq})
\begin{eqnarray} \label{Qeqlin} d\Delta_t=[-\lambda \, dt+ \gamma\, (1-2Q_t)\,
  dB_t]\Delta_t.
\end{eqnarray}     
We claim that any solution $\Delta_t$ of (\ref{Qeqlin}) with positive initial
condition stays positive forever and converges almost surely to $0$.  The
convergence is exponential, and for small $ \sigma$ the rate is twice the
typical measurement time $\tau_{\rm meas}=\gamma^{-2}$. Moreover $\Delta_t$ is a
super-martingale.

{\it Sketch of the proof:} We define $M_t$ by writing $\Delta_t =: \Delta_0 e^{-\lambda\, t} M_t$.
Then $M_0=1$ and $dM_t= \gamma\, M_t(1-2Q_t)\, dB_t$. So $M_t$ is a local
martingale, and It\^o's formula yields immediately that
\[
M_t =\exp \left(-\frac{\gamma^2}{2} \int_0^t (1-2Q_s)^2\,ds + \gamma\int_0^t
  (1-2Q_s)\, dB_s\right).
\]
As $Q_s \in [0,1]$ for all $s$, the Riemann and It\^o integrals on the
right-hand side are finite for every $t$ for almost every sample. Moreover, the
bound $e^{\frac{\gamma^2}{2} \int_0^t (1-2Q_s)^2\,ds} \leq
e^{\frac{\gamma^2}{2}\, s}$ implies that $M_t$ satisfies the Novikov criterion
(see e.g. \cite{Oksen,Kal}), hence that $M_t$ is
well-defined, positive for every $t$, and is a martingale. Consequently, if
$\Delta_0 >0$, $\Delta_t$ is a super-martingale. A na\"{\i}ve scaling argument
indicates that $\int_0^t (1-2Q_s)^2\,ds$ scales like $t$ for large $t$ whereas
$\int_0^t (1-2Q_s)\, dB_s$ scales like $\sqrt{t}$, so that $\lim_{t \to +\infty}
M_t =0$ almost surely, and thus $\lim_{t \to +\infty} \Delta_t =0$ almost surely
as well. One can be a bit more precise. For large $\gamma$ (i.e. $\sigma$ small)
the trajectory $Q_s$ spends most of its time close to $0$ or $1$. Actually, most
of the time $Q_s \propto \sigma$ or $1-Q_s \propto \sigma$, so that
$(1-2Q_s)^2-1=-4Q_s(1-Q_s)$ is most of the time of order $\sim \sigma$. We infer
that $\log M_t =-\frac{\gamma^2}{2} t + \gamma \tilde{B}_t +O(1) $ for large
$\gamma$, where $\tilde{B}_t := \int_0^t \text{sign} (1-2Q_s) \, dB_s$ is a
standard Brownian motion. In particular, the rate at which trajectories
corresponding to different initial conditions but the same measurement outcomes
approach each other is twice the typical measurement time $\tau_{\rm
  meas}=\gamma^{-2}$ for large $\gamma$.

\section{Generalisations and outlook.}

The above study can be generalised to a system in contact with thermal
reservoirs and under continuous time measurement with higher dimensional Hilbert
spaces both for the system and for the probes. Let us briefly present the
general picture. As above we can restrict ourselves to diagonal density matrices
if the Hamiltonian and thermal Lindbladian flows commute. Let
\[ \rho= \sum_\alpha Q_{\alpha}\ \ket{\alpha}\bra{\alpha},\quad \sum_\alpha
Q_\alpha=1,\] be the system density matrix, diagonal in the energy eigen-state
basis. The time evolution of its components is going to be the sum of the
thermal and noisy evolutions, induced by the measurement,
\[ dQ_\alpha = (dQ_\alpha)_{\rm therm} + (dQ_\alpha)_{\rm meas}.\]

As before, we may model the thermal evolution by a Lindbladian \cite{Lind76}.
That is $(d\rho)_{\rm therm}= L_{\rm therm}(\rho)\, dt$, with $L_{\rm
  therm}(\rho) = \sum_a\big[ C_a\rho\, C^\dag_a -\half \{C^\dag_a
C_a,\rho\}\big],$ for some operators $C_a$ acting on ${\cal H}_s$. When
projected on diagonal density matrices, this becomes
\[ (dQ_\alpha)_{\rm therm} = \sum_\gamma\big(
L_{\alpha\gamma}-\ell_\alpha\,\delta_{\alpha;\gamma}\big)Q_\gamma\, dt,\] with
$L_{\alpha\gamma}:=\sum_a |\bra{\alpha}C_a\ket{\gamma}|^2$ and
$\ell_\alpha:=\sum_a\bra{\alpha}C^\dag_a C_a\ket{\alpha}$. The matrix $L$ is
real but (a priori) not symmetric and $\sum_\alpha
L_{\alpha\gamma}=\ell_{\gamma}$ as it should be by compatibility with
$\sum_\alpha Q_\alpha=1$. Demanding that the Gibbs state is a steady state
imposes $\sum_{\gamma}L_{\alpha\gamma} p_\gamma= \ell_\alpha\, p_\alpha$, where
$p_\alpha=e^{-\beta_\alpha}/(\sum_\gamma e^{-\beta_\gamma})$ is the Boltzmann
weight. To rephrase these observations: the matrix
$L_{\alpha\gamma}-\ell_\alpha\,\delta_{\alpha;\gamma}$ describing the thermal
evolution is the infinitesimal generator of a finite state continuous time
Markov process, to which all standard results of probability theory apply.

The evolution equations for continuous time energy measurement have been
described in \cite{Bar,Bar2}. We borrow the notations from \cite{BBB12} which deals
with the simpler non-demolition case. Let us index by $i$, from $1$ to ${\rm
  dim}{\cal H}_p$, the probe measurement outputs. In the diffusive case, these
equations reads $(dQ_{\alpha})_{\rm meas}=Q_{\alpha}\sum_i
[\Gamma(i|\alpha)-\langle{\Gamma(i)}\rangle]dB_i$, with $B_i$ Brownian motions,
$\sum_i B_i=0$, with quadratic variation
$dB_idB_j=[\delta^{ij}p_0(i)-p_0(j)p_0(j)]dt$ where the $p_0(i)$'s are
probabilities on the set of probe measurement outputs, $\sum_i p_0(i)=1$. Here
$\Gamma(i|\alpha)$ are parameters coding for the interaction involved in weak
measurements, $\sum_i p_0(i)\Gamma(i|\alpha)=0$, and
$\vev{\Gamma(i)}=\sum_\alpha \Gamma(i|\alpha)Q_\alpha$. Defining $dW_\alpha:=
\sum_i \Gamma(i|\alpha)dB_i$, these equations becomes
\[(dQ_{\alpha})_{\rm meas}= Q_{\alpha}\big[ dW_\alpha - (\sum_\gamma
Q_{\gamma}dW_\gamma)\big], \quad {\rm with}\ dW_\alpha\, dW_\beta= \bar
\Gamma_{\alpha\beta}\,dt,\] where $\bar \Gamma_{\alpha\beta}= \sum_j p_0(j)
\Gamma(j|\alpha)\Gamma(j|\beta)$. It is compatible with $\sum_\alpha
Q_\alpha=1$.

Two special cases are of particular interest: (a) the probe Hilbert space is of
dimension $2$ as above, there is then only one Brownian motion involved --
because the measurement outputs are in one-to-one correspondence with random
walks -- and we may write $dW_\alpha=\gamma_\alpha dB_t$ for some parameters
$\gamma_\alpha$; and (b) the dimension of the probe Hilbert space is larger than
that of the system Hilbert space so that all Brownian motions $W_\alpha$ are
independent. In both cases we are actually dealing with particular random
processes on probability measures.

These cases will be analysed in \cite{BBnext}, but the general picture is clear.
The invariant measure is going to be localised around unstable `fixed' points,
$Q_\alpha\simeq\delta_{\alpha;\gamma}+{\rm corrections}$, in one-to-one
correspondence with the energy eigen-states. Any given quantum trajectory is
going to wait long periods of time around these points but jump randomly from
time to time to another basin of attraction. They hence yield a fuzzy trajectory
realisation of the thermal Markov chain.  The waiting times are encoded in the
thermal Lindbladian, which here reduces to the Markov matrix $(
L_{\alpha\gamma}-\ell_\alpha\delta_{\alpha;\gamma})$. Indeed, using a frequency
interpretation of the $Q$'s, the off-diagonal elements of $L_{\alpha\gamma}$
codes for a transfer of population from state $\gamma$ to state $\alpha$, and
the natural time associated to transfer from state $\gamma$ to any other states
is $T_\gamma$ with $T_\gamma^{-1} := \sum_{\alpha\not=\gamma} L_{\alpha\gamma}$.
The transition times are going to be determined by the time scale involved in
the measurement process, up to logarithmic corrections. In the discrete
framework these are given by relevant relative entropies \cite{BB11} and are
therefore asymmetric (i.e. the collapse rate of $Q_\gamma$ while converging to
state $\alpha$ is different from the collapse rate of $Q_\alpha$ while
converging to $\gamma$). In the continuous time limit these rates become
$\tau_{\alpha\gamma}$, with $\tau_{\alpha\gamma}^{-1} :=
\half\big(\bar\Gamma_{\alpha\alpha}-2\bar\Gamma_{\alpha\gamma}+\bar\Gamma_{\gamma\gamma}\big)$,
and they are symmetric. These are naturally interpreted as the jumping times
from state $\alpha$ to state $\gamma$, up to logarithmic corrections.

Finally, generalisations to out-of-equilibrium contexts, in which a system is in
contact with different reservoirs at different temperatures, are particularly
interesting. Describing quantum trajectory behaviours in such situations is a
question we plan to address in~\cite{BBnext}.

\appendix
\section{Details about the computation of $\tilde \tau$.}

\label{app:details}

This Appendix is devoted to the proof that the approximate procedure we used to
compute $\tilde{\tau}$ is correct. It is a slight elaboration of an argument
provided to us by an anonymous referee.

For the rest of this appendix, we set $Q_i=p\sigma$ and $Q_f=1-(1-p)\sigma$ and
define
\[ l(Q):=\frac{
  \big(s(Q)-s(Q_i)\big)\big(s(Q_f)-s(Q)\big)}{s'(Q)
  \big(s(Q_f)-s(Q_i)\big)}, \quad 
l_0(Q):=\frac{(Q-Q_i)(Q_f-Q_i)}{(Q_f-Q_i)}\] and set 
\[ I[a,b]:=\int_{a}^{b} \frac{dQ}{g^2(Q)} l(Q) , \quad
I_0[a,b]:=\int_{a}^{b} \frac{dQ}{g^2(Q)} l_0(Q).\]

Recall that general arguments imply that the mean transition time from
$Q_i$ to $Q_f$ is $\tilde{\tau} = 2\gamma^{-2}I[Q_i,Q_f]$.
Our interest is in the asymptotics of $I[Q_i,Q_f]$ in the limit $\sigma\to 0$.
What we really did to estimate $\tilde \tau$ in the main text was some kind of
inversion of limits: we computed the
behaviour of $2\gamma^{-2}I_0[Q_i,Q_f]$ when $\sigma\to 0$ (because for $Q$ in a $\sigma$-independent interval, $s(Q)$ becomes
an affine function of $Q$ in the limit when $\sigma \to 0$). 

Our aim is now to prove that the difference $I[Q_i,Q_f]-I_0[Q_i,Q_f]$ goes to a
finite limit as $\sigma\to 0$.

To compare $I[Q_i,Q_f]$ and $I_0[Q_i,Q_f]$ in the
limit $\sigma\to 0$ we introduce an auxiliary function $M(\sigma)$ such that
$\lim_{\sigma\to 0} M(\sigma) =+\infty$ but $\lim_{\sigma\to 0} M(\sigma)
\sigma \log \sigma=0$ and let $\hat{Q}_i:=p\sigma M(\sigma)$ and
$\hat{Q}_f:=1-(1-p)\sigma M(\sigma)$. We split the integral
\[ I[Q_i,Q_f]=I[Q_i,\hat{Q}_i]+I[\hat{Q}_i,\hat{Q}_f]+I[\hat{Q}_f,Q_f],\]
and similarly for $I_0$. 

We first work on the intermediate integral, and then turn to the first (the
third is analogous). 

\underline{The integral $I[\hat{Q}_i,\hat{Q}_f]$}

We need some preliminary estimates. We define $d(Q,Q'):= \frac{s(Q)-s(Q')}{Q-Q'}$.

{\it Claims :} \\
(a) There is a constant $K$ independant of $Q$ and $\sigma$ such
that $|s'(Q)| \leq K$ for $Q\in [Q_i,Q_f]$. We write this as
$s'(Q)=O_u(1)$ for $Q\in [Q_i,Q_f]$ \footnote{Here and the sequel, it
  is  understood that $O_u$'s, where the subscipt $u$ stands for {\it uniform}, are $Q$-independant in the specified
  interval.}. For $Q\in [\hat{Q}_i,\hat{Q}_f]$,  
 $s'(Q)=1+O_u(1/M(\sigma))$ \\
(b) For $Q\in [\hat{Q}_i,\hat{Q}_f]$ one has
$d(Q,Q_i) =1+O_u(1/\sqrt{M(\sigma)})$ and $d(Q_f,Q)  =1+O_u(1/\sqrt{M(\sigma)})$. \\
(c) The ratio $\frac{s(Q_f)-s(Q_i)}{Q_f-Q_i}$ differs
from $1$ by an $O_u(\sqrt{\sigma})$.\\ 
(d)  The ratio $\frac{l(Q)}{l_0(Q)}$ differs
from $1$ by an $O_u(\sqrt{\sigma})$.

{\it Proof of (a):} Recall that $s'(Q) = e^{\sigma \left(\frac{p}{Q}+
    \frac{1-p}{1-Q}+(2p-1) \log \frac{1-Q}{Q}\right)}$. The derivative of the
funtion in parenthesis is $\frac{Q-p}{g^2(Q)}$, so to get the uniform bound it
suffices to compute $s'(Q)$ at $Q=p$ and at the two boundaries. Now
$s'(p)=1+O_u(\sigma \log M(\sigma))$. Comparing to the boundary values leads to
$s'(Q)=O_u(1)$ for $Q\in [Q_i,Q_f]$ and $s'(Q)=1+O_u(1/M(\sigma))$ for $Q\in
[\hat{Q}_i,\hat{Q}_f]$.

{\it Proof of (b):} The trick is to introduce another function $N(\sigma)$ such
that $\lim_{\sigma\to 0} N(\sigma) =+\infty$ but $N(\sigma)=o(M(\sigma))$. Let
$\bar{Q}_i:= p\sigma N(\sigma)$ and write
$s(Q)-s(Q_i)=\left(s(Q)-s(\bar{Q}_i)\right)+\left(s(\bar{Q}_i)-s(Q_i)\right)$.
The intermediate value theorem ensures the existence of some $Q_> \in
]\bar{Q}_i,\hat{Q}_f[$ such that $s(Q)-s(\bar{Q}_i) = (Q-\bar{Q}_i)s'(Q_>)$ and
some $Q_< \in ]Q_i, \bar{Q}_i[$ such that $s(\bar{Q}_i)-s(Q_i)=
(\bar{Q}_i-Q_i)s'(Q_<)$. Then
\[ \frac{s(Q)-s(Q_i)}{Q-Q_i}= s'(Q_>) +
\frac{(\bar{Q}_i-Q_i)}{Q-Q_i}(s'(Q_<)-s'(Q_>)).\] Claim (a) but
applied for $N$ instead of $M$ implies that the first term is
$1+O_u(1/N(\sigma))$. Claim (a) again implies that
$s'(Q_<)-s'(Q_>)=O_u(1)$. As $1/(Q-Q_i)$ is maximal in the interval under
consideration at $Q=\hat{Q}_i$ the second term is
$O_u(N(\sigma)/M(\sigma))$. Balancing the two terms leads to choose $N(\sigma)
\propto \sqrt{M(\sigma)}$, proving the claim. The second claim is proved
analogously. 

{\it Proof of (c):} The trick is to introduce a new auxiliary function
$\tilde{N}(\sigma)$.  This time we split $s(Q_f)-s(Q_i)$ in three pieces as
$s(Q_f)-s(1-(1-p)\sigma \tilde{N}(\sigma))+ (s(1-(1-p)\sigma \tilde{N}
(\sigma))-s(p\sigma \tilde{N}(\sigma)))+ (s(p\sigma \tilde{N}(\sigma))-s(Q_i))$.
The proof proceeds as the proof in the previous claim. This time the optimum is
to choose $\tilde{N}(\sigma) \propto \sqrt{\sigma}$. We leave the details to the
reader.

{\it Proof of (d):} Observe that
\[ \frac{l(Q)}{l_0(Q)}=\frac{d(Q,Q_i)d(Q_f,Q)}{s'(Q)
  d(Q_f,Q_i)}= 1+O_u(1/\sqrt{M(\sigma)}).\]
All the factors are close to $1$ with deviations bounded, according to the
previous claims, by
$O_u(1/M(\sigma))$, $O_u(1/\sqrt{M(\sigma)})$ or $O_u(\sqrt{\sigma)})$. The
softest  one is $O_u(1/\sqrt{M(\sigma)})$, which proves the claim.  

\vspace{.4cm}

This leads to our first goal.

\underline{{\it Claim:}} \\
The difference  $I[\hat{Q}_i,\hat{Q}_f]-I_0[\hat{Q}_i,\hat{Q}_f]$ is $O(\frac{\log
  \sigma}{\sqrt{M(\sigma)}})$.

{\it Proof:} Recall that $I[\hat{Q}_i,\hat{Q}_f]=\int_{\hat{Q}_i}^{\hat{Q}_f} \frac{dQ}{g^2(Q)}
l(Q)$ (and the analog relating  $I_0$ and $l_0(Q)$) so that we need to bound 
\[\int_{\hat{Q}_i}^{\hat{Q}_f} \frac{dQ}{g^2(Q)}
(l(Q)-l_0(Q))=\int_{\hat{Q}_i}^{\hat{Q}_f}
\frac{dQ}{g^2(Q)} \left(\frac{l(Q)}{l_0(Q)}-1\right)l_0(Q).\]
By the previous claim, $\frac{l(Q)}{l_0(Q)}-1=O_u(1/\sqrt{M(\sigma)})$ which by
uniformity can be taken outside the integral. The remaining integral is nothing
but $I_0[\hat{Q}_i,\hat{Q}_f]$ which we know is $\sim \log
\sigma$, and the result follows. 

\underline{The integral $I[Q_i,\hat{Q}_i]$}

Let $\mM >0$ be a constant. We start with the evaluation of some integrals, our
main interest being their large $\mM$ behavior. The
first observation is that:
\[ J_0:=\int_{1}^{\mM} \frac{dY}{Y^2} (Y-1) =\log \mM -1+\frac{1}{\mM}=\log \mM
-1+ O(1/\mM).\] Taking $Y=Q/Q_i$, we note that $\int_{Q_i}^{\hat{Q}_i}
\frac{dQ}{Q^2} (Q-Q_i)={J_0}_{|\mM = M(\sigma)}$.

Second we consider 
\[J:=\int_{1}^{\mM} \frac{dY}{Y^2} \int_{1}^{Y} dy \, e^{1/y-1/Y},\] which we
expand at large $\mM$. Interchanging integrals
\[ J =\int_{1}^{\mM} dy \int_y^{\mM} \frac{dY}{Y^2}e^{1/y-1/Y} = \int_{1}^{\mM}
dy \left( e^{1/y-1/\mM}-1\right)= 1-\mM+e^{-1/\mM} \int_{1}^{\mM} dy \,
e^{1/y}. \] Now using $e^{1/y} = \left(e^{1/y}-1-1/y \right)+ \left(1+1/y
\right) $ we get 
\[\int_{1}^{\mM} dy \, e^{1/y}=\mM -1+\log \mM + \int_{1}^{\infty} dy \, \left(e^{1/y}-1-1/y
\right)+O(1/\mM).\] Setting $\iI := \int_{1}^{\infty} dy \, \left(e^{1/y}-1-1/y
\right)$, a straightforward expansion yields
\[ J=-1+ \iI + \log \mM -\frac{ \log \mM }{\mM} + O(1/\mM).\]
In particular 
\[ J-J_0 = \iI -\frac{ \log \mM }{\mM} + O(1/\mM)\] 

We need some estimates again.

{\it Claims :} \\
(e) For $Q\in [Q_i, \hat{Q}_i]$ we have $s(Q)-s(Q_i)=O_u(\sigma M(\sigma))$.\\
(f) For $Q\in [Q_i, \hat{Q}_i]$ we have
$\frac{s(Q_f)-s(Q)}{s(Q_f)-s(Q_i)}=1+O_u(\sigma M(\sigma))$.\\ 
(g) For $Q\in [Q_i, \hat{Q}_i]$ we have $Q^2/g(Q)^2=1+O_u(\sigma M(\sigma)).$\\
(h) $I[Q_i,\hat{Q}_i]=(1+O(\sigma M(\sigma)))J_{|\mM = M(\sigma)} $.

{\it Proof of (e):} We write $s(Q)-s(Q_i)=(Q-Q_i)s'(Q_<)$ for some $Q_< \in
]Q_i, \hat{Q}_i[$. Note that $|Q-Q_i|\leq \hat{Q}_i-Q_i=O_u(\sigma M(\sigma))$.
By claim (a) $s'(Q_<)=O_u(1)$ because $[Q_i, \hat{Q}_i] \subset [Q_i,Q_f]$.

{\it Proof of (f):} We write
$\frac{s(Q_f)-s(Q)}{s(Q_f)-s(Q_i)}=1-\frac{s(Q)-s(Q_i)}{s(Q_f)-s(Q_i)}$ and use
claims (e) and (c). 

{\it Proof of (g):} Obvious and left to the reader.

{\it Proof of (h):} From (f) and (g) we infer that
\[I[Q_i,\hat{Q}_i] = \int_{Q_i}^{\hat{Q}_i} \frac{dQ}{Q^2}
\frac{s(Q)-s(Q_i)}{s'(Q)} (1+O_u(\sigma M(\sigma))) =(1+O(\sigma
M(\sigma)))\int_{Q_i}^{\hat{Q}_i} \frac{dQ}{Q^2} \frac{s(Q)-s(Q_i)}{s'(Q)}. \]
The last integral is
\[\int_{Q_i}^{\hat{Q}_i} \frac{dQ}{Q^2}\int_{Q_i}^Q dq \,
\frac{s'(q)}{s'(Q)}=\int_1^{M(\sigma)} \frac{dY}{Y^2} \int_1^Y dy e^{1/y-1/Y}
e^{k(\sigma,y,Y)}\] where \[k(\sigma,y,Y):=\sigma \left( \frac{1-p}{1-p\sigma
    y}-\frac{1-p}{1-p\sigma Y}+(2p-1)\log \frac{Y(1-p\sigma y)}{y(1-p\sigma Y)}
\right),\] as is seen by the dilation $Q=p\sigma Y$, $q=p\sigma y$. In the
interval $[Q_i,\hat{Q}_i]$, note that $k(\sigma,y,Y)=O_u(\sigma \log
M(\sigma))$, so that 
\[\int_{Q_i}^{\hat{Q}_i} \frac{dQ}{Q^2} \frac{s(Q)-s(Q_i)}{s'(Q)}= (1+O(\sigma \log
M(\sigma))\int_1^{M(\sigma)} \frac{dY}{Y^2} \int_1^Y dy \, e^{1/y-1/Y}\]
The last integral is nothing but $J$ and the result follows.

By a parallel but much simpler argument, we would get that
$I_0[Q_i,\hat{Q}_i]=(1+O(\sigma M(\sigma))) {J_0}_{|\mM = M(\sigma)}$. 

\vspace{.4cm}

This leads to our second goal:

\underline{{\it Claim:}} \\
The difference $I[Q_i,\hat{Q}_i]-I_0[Q_i,\hat{Q}_i]$ is equal to
$\iI +O\left(\frac{\log
  M(\sigma)}{M(\sigma)}\right) + O(\sigma M(\sigma)\log M(\sigma))$

{\it Proof:} Using (h) we know that \begin{eqnarray*}
  I[Q_i,\hat{Q}_i]-I_0[Q_i,\hat{Q}_i]&=&(1+O(\sigma M(\sigma))) J_{|\mM = M(\sigma)}-(1+O(\sigma M(\sigma))) {J_0}_{|\mM = M(\sigma)} \\
  & = & (J-J_0)_{|\mM = M(\sigma)}+ J_{|\mM = M(\sigma)}O(\sigma \log
  M(\sigma))-{J_0}_{|\mM = M(\sigma)}O(\sigma M(\sigma)).\end{eqnarray*} We know
from the preliminary computations that $J-J_0=\iI +O\left(\frac{ \log \mM }{\mM}\right)$ and that both
$J$ and $J_0$ are $O(\log \mM)$, which yields
\[ I[Q_i,\hat{Q}_i]-I_0[Q_i,\hat{Q}_i] = \iI +O\left(\frac{ \log M(\sigma) }{
    M(\sigma)}\right) +O(\sigma M(\sigma)\log M(\sigma)),\] proving the claim.

Of course, the same formula holds for the third integral. Putting the three
pieces together we have proved that \[I[Q_i,Q_f]=I_0[Q_i,Q_f]+2\iI
+O\left(\frac{ \log M(\sigma) }{ M(\sigma)}\right) + O(\sigma M(\sigma)\log
M(\sigma)) + O\left(\frac{\log \sigma}{\sqrt{M(\sigma)}}\right).\] 
The third term is always softer than the first. 
Balancing the two
remaining terms leads to take
$M(\sigma)=\sigma^ {-2/3}$ proving that 
\[ I[Q_i,Q_f]-I_0[Q_i,Q_f]=2\iI
+O(\sigma^{1/3} \log \sigma).\] We know that 
\[I_0[Q_i,Q_f] = \Big( \frac{Q_i+Q_f-2Q_iQ_f}{Q_f-Q_i}\log
\frac{Q_f(1-Q_i)}{Q_i(1-Q_f)}-2\Big) \underset{\sigma\to 0}{\sim} -2\log \sigma
-2-\log p(1-p) +O(\sigma).\]
We recall that $Q_i=\sigma p$ , $Q_f=1-\sigma(1-p)$. Hence we have proved that
\begin{equation} \label{eq:final-result} \frac{\gamma^2}{2} \tilde \tau =
  I[Q_i,Q_f]\underset{\sigma\to 0}{=} -2\log \sigma +2(\iI -1) -\log
  p(1-p)+O(\sigma^{1/3} \log \sigma).
\end{equation}

\bigskip

\noindent {\it Acknowledgements}: This work was in part supported by ANR
contract ANR-2010-BLANC-0414. We thank Antoine Tilloy for discussions.  We would
like to thank the anonymous referee for his useful report and contribution
to Appendix A.

\end{document}